\documentclass[preprint,12pt]{elsarticle}

\usepackage{amssymb}

\newcommand{\be}{\begin{equation}}
\newcommand{\ee}{\end{equation}}

\begin{document}

\begin{frontmatter}



\title{A symmetry of the spatially flat Friedmann equations with 
barotropic fluids}


\author{Valerio Faraoni}

\address{Physics Department and STAR Research Cluster, Bishop's 
University, 2600 College Street, Sherbrooke, Qu\'ebec, Canada 
J1M~1Z7}

\begin{abstract}
A string-inspired duality symmetry of the spatially flat 
Friedmann equations of general-relativistic cosmology is 
discussed and generalized, providing a map between exact 
solutions 
corresponding to different values of the 
barotropic index. 
\end{abstract}


\end{frontmatter}



\section{Introduction}

In modern theoretical physics, the symmetry group of a theory 
plays a crucial role and one of the first questions asked of a 
new theory is which are its symmetries. Here we point out a 
symmetry property of the Einstein-Friedmann equations, the 
fundamental equations of  
spatially homogeneous and isotropic 
Friedmann-Lema\^itre-Robertson-Walker (FLRW) cosmology, for the 
case of flat spatial sections. A special 
subcase has been discussed extensively in the literature 
\cite{Chimento02, Aguirregabiriaetal03, ChimentoLazkoz03, 
DabrowskiStachowiakSzydlowski03, AguirregabiriaChimentoLazkoz04, 
Calcagni05, SzydlowskiGodlowskiWojtak06,  
ChimentoLazkoz06, ChimentoZimdhal06, ChimentoPavon06, 
DabrowskiKieferSandhoefer06, 
CaiLiPiaoZhang07, Chimentoetal07, CataldoChimento08, 
Capozzielloetal09,   
WangYang09, CapozzielloFaraonibook, 
Caietal10, PucheuBellini10, Chimentoetal11}, 
and  it very much resembles a well known  duality of 
string cosmology. In spite of a significant 
amount of literature on the symmetries of the Friedmann 
equations, with particular attention to generating new 
analytical solutions 
from known ones, in both classical and quantum cosmology 
(\cite{Chimento02, Aguirregabiriaetal03, ChimentoLazkoz03, 
DabrowskiStachowiakSzydlowski03, AguirregabiriaChimentoLazkoz04, 
Calcagni05, SzydlowskiGodlowskiWojtak06,  
ChimentoLazkoz06, ChimentoZimdhal06, ChimentoPavon06, 
DabrowskiKieferSandhoefer06, 
CaiLiPiaoZhang07, Chimentoetal07, CataldoChimento08, 
Capozzielloetal09,  
WangYang09, CapozzielloFaraonibook, 
Caietal10, PucheuBellini10, Chimentoetal11} and 
references therein), to the best of our knowledge the general 
symmetry  has not been commented upon\footnote{See also 
\cite{Rosu} for a solution of the Einstein-Friedmann equations 
(which, for a barotropic fluid, reduce to a well known Riccati 
equation \cite{AmJP}) using methods of supersymmetric quantum 
mechanics.}.

We restrict ourselves to spatially flat  FLRW cosmology and to 
the line element
\be
ds^2=-dt^2+a^2(t) \left( dx^2+dy^2+dz^2 \right)
\ee
in comoving coordinates $\left( t,x,y,z \right)$. The scale 
factor is taken to be dimensionless, while the coordinates 
carry the dimension of a length. We assume a 
perfect fluid as the matter source in the Einstein equations of 
general relativity, which reduce to \cite{LandauLifschitz, Wald, 
Liddle}
\begin{eqnarray}
&& \frac{ \ddot{a} }{a}  =  -\, \frac{4\pi }{3} \left( \rho +3P 
\right) \,, \label{2}\\
&&\nonumber\\
&& \left( \frac{ \dot{a} }{a} \right)^2  =   \frac{8\pi }{3} 
\, \rho  \,, \label{3}
\end{eqnarray}
where  $\rho(t) $ and $P(t)$ are the  energy density and pressure 
of the perfect fluid, respectively, and we use units in which 
Newton's constant $G$ and the speed of light are unity.  
The perfect fluid is described by the stress-energy tensor 
\be
T_{ab}=\left( P+\rho \right) u_a u_b +Pg_{ab} \,,
\ee
where $u^a$ is the fluid (timelike) four-velocity which 
has components $u^{\mu}=\delta^{\mu}_0$ in comoving coordinates. 
We further assume that the perfect fluid is ruled by the barotropic 
equation of state 
\be \label{4}
P=w\rho \equiv \left( \gamma-1 \right) \rho \,,
\ee
where the barotropic index $\gamma$ is constant and $w$ 
is usually referred 
to as the equation of state parameter. As is well known, the 
covariant conservation equation $\nabla^b T_{ab}=0$ yields
\be \label{5}
\dot{\rho}+3H \left( P+\rho \right)=0 \,,
\ee
where $H\equiv \dot{a}/a$ is the Hubble parameter. This 
conservation equation can also be obtained directly from 
eqs.~(\ref{2}) and~(\ref{3}) and is immediately integrated 
yielding $\rho$ as a function of the scale factor,
\be \label{6}
\rho (a) =\frac{ \rho_0 }{a^{3\gamma} } =\frac{\rho_0}{ 
a^{3(w+1)}} 
\,.
\ee

\section{A symmetry of the Einstein-Friedmann equations}

Let us perform the operation
\be  
a \rightarrow \alpha \equiv \frac{1}{a}  \,;\label{7}
\ee
then $H=-\dot{\alpha}/\alpha$, while $ \frac{ \ddot{a} 
}{a}=2\left( 
\frac{ \dot{\alpha} }{\alpha} \right)^2 
-\frac{\ddot{\alpha} }{\alpha}$  and eqs.~(\ref{2}) and (\ref{3}) 
become
\begin{eqnarray}
&& \frac{ \ddot{\alpha}}{\alpha} =  -\, \frac{4\pi }{3}  
\left(3\tilde{\gamma}-2  \right) \rho  \,, \label{8}\\
&&\nonumber\\
&& \left( \frac{ \dot{\alpha}}{\alpha} \right)^2 = 
\frac{8\pi}{3} \, \rho  \,, \label{9}
\end{eqnarray}
where $\tilde{\gamma} \equiv -\gamma$. That is, eqs.~(\ref{2}) 
and~(\ref{3}) are invariant in form under the transformation
\be \label{11}
a \rightarrow \alpha \equiv \frac{1}{a} \,, \;\;\;\;\;\;\;\;
\gamma \rightarrow \tilde{\gamma} =-\gamma  \,.
\ee
The 
inversion of the scale factor can be compensated by the change in 
the equation of state 
\be
P=w\rho \rightarrow P=\tilde{w}\rho=-\left( w+2 \right)\rho 
\label{12}
\ee
where $\tilde{w}=-(w+2)$ (equivalent to 
$\tilde{\gamma}=-\gamma$). Therefore, there is a duality in the 
equations ruling spatially flat FLRW cosmology which maps 
expanding universes into contracting ones, and {\em vice-versa}. 
Assuming that the barotropic index $\gamma \neq 0$ and that the 
weak energy condition 
($\rho \geq  0$ and $\rho+P =\gamma \rho \geq 0$) \cite{Wald}  is 
satisfied by 
a universe with scale factor $a(t)$ and perfect fluid equation of 
state~(\ref{4}), the ``dual universe'' with scale factor 
$\alpha(t)=1/a(t)$ certainly violates it since $\tilde{\gamma} 
\rho=-\gamma \rho < 0$. The perfect fluid associated with the 
dual universe is, for sure, a phantom fluid. For example, the 
dual 
of a stiff fluid universe with $P=\rho$ is a strongly phantom 
fluid with $P=-3\rho$; this situation can be realized by a 
massless scalar field with no potential. In fact, the energy 
density and pressure of a minimally coupled scalar are
\be
\rho=\frac{ \dot{\phi}^2}{2}+V(\phi)
\ee
and 
\be
P=\frac{ \dot{\phi}^2}{2}-V(\phi) \,;
\ee 
setting $V\equiv 0$ yields 
$P=\rho$ and it is well known that in the FLRW space a minimally 
coupled  scalar 
field  can be given a fluid description \cite{Madsen}.

The conservation equation~(\ref{5}), of course, does not change 
by 
virtue of being a consequence of the field equations and not an 
independent equation. It is straightforward to check directly, 
using eqs.~(\ref{8}) and (\ref{9}), that 
\be \label{13}
\dot{\rho}+3\, \frac{\dot{\alpha}}{\alpha}\, \tilde{\gamma}\rho=0
\ee
the solution of which is, of course, 
$\rho(\alpha)=\rho_0/\alpha^{3\gamma}$ and can be obtained 
directly from eq.~(\ref{6}).

For $\gamma=$const.$\neq 0$, the solution of eqs.~(\ref{2}) and 
(\ref{3}) is the power-law 
\be \label{powerlaw}
a(t)=a_0 \left| t-t_* \right|^{\frac{1}{3\gamma}} =
a_0 \left| t-t_* \right|^{\frac{1}{3( w+1)}}
\ee
as is well known. For $\gamma>0$  and $t>t_*$ this is a Big Bang universe 
with initial singularity at $t_*$. The ``dual'' solution\footnote{Since the duality transformation 
does not involve explicitly the time $t$, it acts only in one of the two branches $t<t_*$ and 
$t>t_*$.} 
is given 
by 
\be
\alpha(t)= \frac{  \alpha_0}{ 
\left| t-t_* \right|^{\frac{1}{3\gamma}} }
\ee
and is a contracting universe beginning from an $a=\infty$ singularity at $t_*$,  as a consequence 
of the fact that the fluid with 
$P+\rho=\gamma \rho $ has changed into a phantom fluid with 
$ \tilde{\gamma}\rho= - \gamma \rho  $ in the dual solution.
For $\gamma>0$  and $t< t_*$  there is a duality between a contracting universe ending in a Big 
Crunch singularity at $t_*$ and an expanding superaccelerating ({\em i.e.}, $\dot{H}>0$) universe 
ending in a 
Big Rip singularity with $a=\infty$ at $t_*$.

For $\gamma=0$, the solution of eqs.~(\ref{2}) and (\ref{3}) is 
the expanding de Sitter space $a(t)=a_0 \, \mbox{e}^{Ht}$ 
($H=$const.$>0$) and the ``dual'' universe is the contracting de 
Sitter space $\alpha(t)=\alpha_0 \, \mbox{e}^{-Ht}$. Minkowski 
space (the degenerate case with $H=0$ and $\rho=P=0$) is 
obviously a fixed point of the transformation~(\ref{11}), but 
a trivial one.

The duality symmetry has been extended to brane world models  
with bulk effects switched off in \cite{ChimentoLazkoz03} (see 
also \cite{NojiriOdintsov03}).

\section{Generalizing the duality}

Since the Einstein-Friedmann equations are so fundamental and in 
view of the extensive applications of the duality symmetry 
introduced in the previous section to classical and quantum 
cosmology 
\cite{Chimento02, Aguirregabiriaetal03, ChimentoLazkoz03, 
DabrowskiStachowiakSzydlowski03, AguirregabiriaChimentoLazkoz04, 
Calcagni05, SzydlowskiGodlowskiWojtak06,  
ChimentoLazkoz06, ChimentoZimdhal06, ChimentoPavon06, 
DabrowskiKieferSandhoefer06, 
CaiLiPiaoZhang07, Chimentoetal07, CataldoChimento08, 
Capozzielloetal09,  
WangYang09, CapozzielloFaraonibook, 
Caietal10, PucheuBellini10, Chimentoetal11}, it is useful to 
generalize it by asking  what is required in order to map a 
perfect fluid solution  of eqs.~(\ref{2}) and (\ref{3}) with 
constant barotropic index  $\gamma_1$ into a  solution with 
index $\gamma_2$ (with 
$\gamma_1, \gamma_2\neq 0$). All FLRW metrics 
are conformally flat, and the spatially flat FLRW line element 
is explicitly conformal to the Minkowski one 
$ds_0^2=-d\eta^2+dx^2+dy^2+dz^2$. Using the conformal time $\eta 
$ defined by $dt=a d\eta$ 
we have,\footnote{For the solution given by 
eq.~(\ref{powerlaw}), the conformal time is 
$\eta(t)=\frac{3\gamma}{\left( 3\gamma -1 \right) a_0} \left| 
t-t_*\right|^{\frac{1-3\gamma}{3\gamma}} $ and $ a(\eta)= \left( 
\frac{3\gamma}{3\gamma -1 } \right)^{\frac{1}{3\gamma -1}} a_0 \, 
\eta^{\frac{1}{3\gamma-1}} $.}  for the two FLRW 
solutions, 
\begin{eqnarray}
ds_1^2  & = & -dt^2+a_1^2(t) \left( dx^2+dy^2+dz^2\right) 
=\Omega_1^2 \, ds_0^2 \,,\\
&&\nonumber\\
ds_2^2  & = & -dt^2+a_2^2(t) \left( dx^2+dy^2+dz^2\right) 
=\Omega_2^2 \, ds_0^2 \,,
\end{eqnarray}
where $\Omega_{1,2}=a_{1,2}( \eta) $ is the conformal factor for 
each case, and 
\begin{eqnarray}
a_1(t) & = & a _* \left| t-t_* \right|^{\frac{1}{3\gamma_1}} 
\,,\\
&&\nonumber\\
a_2(t) & = & a_* \left| t-t_* \right|^{\frac{1}{3\gamma_2}} 
\end{eqnarray}
(the constant $a_*$ needs not be the same in both cases, but it 
is irrelevant because it can always be eliminated by rescaling 
the coordinates). Clearly, it is
\be 
ds_2^2=\Omega_2^2 \, ds_0^2 =\Omega^2 ds_1^2 \,,
\ee
where 
\be
\Omega = \frac{\Omega_2}{\Omega_1}= \frac{a_2}{a_1} = 
\mbox{const.} \cdot \left| 
t-t_*\right|^{ \frac{\gamma_2-\gamma_1}{\gamma_1\left( 
3\gamma_2-1\right)}}  
\ee
($\gamma_2 \neq 1/3$). In other words, we  can obtain a solution 
of eqs.~(\ref{2}) and 
(\ref{3}) simply by conformally transforming another 
solution, because they are both  conformal to Minkowski space 
and, therefore, are conformally related. Of course, one could 
simply 
obtain $ \left| t-t_* \right| $ from $a_1$ and substitute it into 
$a_2$ to obtain  $ a_2 = a_1^{\gamma_1/\gamma_2}$, without the 
need to identify conformal transformations, but then the 
discussion  would not be generally covariant.

The transformation $\Gamma_{12}$ described by 
\begin{eqnarray}
&& \gamma_1 \longrightarrow \gamma_2 \;\;\;\;\;\;\;\; 
\left(\gamma_1\neq 
0 \, ,  \;\; \gamma_2\neq 0, \frac{1}{3} \right), \\
&&\nonumber\\
&& a_1(t) \longrightarrow a_2(t)=\left[ 
a_1(t)\right]^{\gamma_1/\gamma_2} \,,
\end{eqnarray}
is a symmetry of the Einstein-Friedmann equations~(\ref{2}) and 
(\ref{3}). 
This property can be checked explicitly by substituting the 
expression $a_2=Aa_1^{\gamma_1/\gamma_2}$ (where $A$ is a 
constant) into eqs.~(\ref{2}) and (\ref{3}). Eqs.~(\ref{2}) and 
(\ref{3}) are satisfied  if, given the energy density  $ \rho_1= 
\rho_1^{(0)} /a_1^{3\gamma_1} $, it is
\be
\rho_1 \longrightarrow \rho_2= \left( \frac{\gamma_1}{\gamma_2} 
\right)^2 \rho_1 \,,
\ee
while 
\be
H_1\equiv \frac{ \dot{a}_1}{a_1} \longrightarrow 
H_2\equiv \frac{ \dot{a}_2}{a_2} =\frac{\gamma_1}{\gamma_2}\, H_1 
\,.
\ee
The transformations $\Gamma_{ij}$ form a non-commutative group 
with respect to the operation of composition of 
transformations~$\circ$.  In fact, $\Gamma_{jk} \circ 
\Gamma_{ij} = \Gamma_{ik}$, the zero element is the identity 
$\Gamma_{ii}$, and the inverse of $\Gamma_{ij}$ is 
$\Gamma_{ij}^{-1}=\Gamma_{ji}$.

As a special case, one recovers the map of the previous section 
which has been studied in the literature
\be
\gamma_1\equiv  \gamma \longrightarrow \gamma_2= -\gamma \,,
\ee
for which $a_2=a_1^{\gamma_1/\gamma_2}=1/a$.

\section{Discussion}
\label{}

The duality $\left( a, \gamma \right) \longleftrightarrow \left( 
a^{-1}, -\gamma \right)$ is used in a variety of applications to 
generate new solutions 
from  old ones and it is of  practical utility. It is interesting 
to try  to understand if this symmetry comes from more 
fundamental 
theories  which leave a memory in the limit to 
general-relativistic cosmology (keeping in 
mind that this is not a symmetry of the full Einstein 
equations, nor the well known time-reversal symmetry). The 
duality discussed here resembles a duality in the 
spatially flat FLRW space of pre-big bang theories \cite{878, 
420, 
859, 435}, which is generalized by another symmetry in  the  
spatially flat FLRW cosmology of vacuum Brans-Dicke theory 
\cite{BD}. Restricting ourselves to the purely 
gravitational sector of the theory, the 
Brans-Dicke field $\phi$ still acts as an effective form of 
matter, 
producing the equations of motion \cite{BD}
\begin{eqnarray}
\dot{H} &=& -\frac{\omega}{2} \left( \frac{ \dot{\phi}}{\phi} 
\right)^2 +2H\, \frac{\dot{\phi}}{\phi} +\frac{1}{2(2\omega+3)} 
\left( \phi \, \frac{dV}{d\phi} -2V \right) \,,\label{17}\\
&&\nonumber\\
H^2 &=& \frac{\omega}{6} \left( \frac{\dot{\phi}}{\phi} \right)^2 
-H \frac{\dot{\phi}}{\phi} +\frac{V(\phi)}{6\phi} \,,\label{18}
\end{eqnarray}
where $V(\phi)$ is the scalar field potential (absent in the 
original Brans-Dicke formulation but naturally present in high 
energy theories) and $\omega $ is the 
Brans-Dicke parameter \cite{BD}. In the absence of matter, the 
Brans-Dicke field satisfies the equation 
\be
\ddot{\phi}+3H\dot{\phi} +\frac{1}{2\omega+3} \left( -\phi\, 
\frac{dV}{d\phi}+2V \right)=0 \,.
\ee
It is well known that vacuum, spatially flat FLRW cosmology 
enjoys a duality symmetry \cite{Lidsey95, Lidsey96}. Let us use new 
variables
\be \label{21}
\beta \equiv \ln a \,, \;\;\;\;\;\;
\Phi \equiv -\ln \left( G\phi \right) 
\ee
instead of $a$ and $\phi$; then the duality transformation takes 
the form \cite{Lidsey95, Lidsey96} 
\begin{eqnarray}
\beta & \rightarrow & \left( \frac{3\omega +2}{3\omega+4} \right) 
\beta -2 \left( \frac{\omega+1}{3\omega+4} \right) \Phi \,, 
\label{22}\\
&&\nonumber\\
\Phi & \rightarrow & \left( -\frac{6 }{3\omega+4} \right) 
\beta - \left( \frac{3\omega+2}{3\omega+4} \right) \Phi \,, 
\label{23}
\end{eqnarray}
for $\omega\neq -4/3$.
This transformation generalizes the duality present in the 
effective action of string theories \cite{878, 420, 859, 435}
\begin{eqnarray}
\beta & \rightarrow &  -\beta \,, \label{24}\\
&&\nonumber\\
\Phi & \rightarrow & \Phi-6\beta \,, \label{25}
\end{eqnarray}
which is reproduced by eqs.~(\ref{22}) and (\ref{23}) for 
$\omega =-1$ (remember that the bosonic string theory reduces to 
a  Brans-Dicke theory with this value of the Brans-Dicke 
parameter \cite{bosonic}). In general relativity, the effective 
gravitational coupling corresponding to $\phi^{-1}$ is a 
constant, not a dynamical variable, and eq.~(\ref{25}) would not 
make sense there; eq.~(\ref{24}) corresponds to the 
transformation~(\ref{7}). However, to make this transformation a 
true symmetry, it is required that a perfect barotropic fluid 
with 
constant equation of state be present, and that the 
transformation $\gamma \rightarrow -\gamma$  of this fluid 
accompanies~(\ref{7}). 
By contrast, string theory enjoys the symmetry~(\ref{24}) 
and~(\ref{25}) {\em in vacuo} (but $\Phi$ acts as a form of 
effective matter). In general, the $\Phi$-field does not behave 
as a perfect fluid with constant equation of state due to 
its dynamical nature, which causes the effective parameter $w$ 
(or $\gamma$) to be time-dependent.  Therefore, in spite of the 
coincidence of 
eqs.~(\ref{7}) and~(\ref{24}), we conclude that there is no 
direct link between the symmetry of spatially flat FLRW space in 
general relativity and the symmetry~(\ref{24}), (\ref{25}) of 
pre-big bang theories. Similarly, in the 
$\omega\rightarrow\infty$ 
limit  of Brans-Dicke theory in which general relativity is  
(usually but not always\footnote{Situations in which this limit 
does not 
reproduce general relativity have been discussed in 
\cite{BDlimit}.}) reproduced, eqs.~(\ref{22}) and~(\ref{23}) 
do not reproduce the symmetry~(\ref{11}).

The duality $\left( a, \gamma \right) \longleftrightarrow \left( 
a^{-1}, -\gamma \right)$ tuns out to be only a special case of 
the more general group of symmetries  $\left( a_1, \gamma_1 
\right) 
\longleftrightarrow \left( a^{\gamma_1/\gamma_2}, \gamma_2  
\right)$; the application of this more general duality to 
classical cosmological scenarios and to the Wheeler-DeWitt 
equation will be the subject of future publications.

This work is supported by the Natural Sciences and  
Engineering Research Council of Canada (NSERC).







\end{document}